\documentclass[]{interact}

\usepackage[T1]{fontenc}% Replaces fontenc when using LuaLaTeX or XeLaTeX
\usepackage{epstopdf}% To incorporate .eps illustrations using PDFLaTeX, etc.
\usepackage{tcolorbox}
\usepackage{libertinus}
\usepackage{libertinust1math}
\usepackage[%
breaklinks=true,
colorlinks=false,
hidelinks,% Keine farbigen Links
bookmarksnumbered=true% Lesezeichen auch nummerieren
]{hyperref}% Sorgt auch dafür, dass im PDF-Dokument die richtige Seitenzahl steht (z. B. I, II am Anfang)

\usepackage[labelfont=bf,font=small]{caption}% Figure 1 in bold (https://tex.stackexchange.com/questions/32459/figure-how-to-have-figure-1-5-in-bold), smaller caption font (font size overview: https://tex.stackexchange.com/questions/24599/what-point-pt-font-size-are-large-etc)

\usepackage{floatrow}
\usepackage{graphicx}
\usepackage{siunitx}
\usepackage{mwe}
\usepackage[style=american]{csquotes}% Use double quotes first which is an alternative style in british english
\usepackage[label font=bf,labelformat=simple,caption=false]{subfig}
\usepackage[final,draft=false]{microtype}% Character protrusion

\usepackage{fancyhdr}
\begin{document}
% Use proportional numbers throughout the document. This does not influence table numbers which are still monospaced
\fontfamily{LinuxLibertineT-LF}\selectfont%(https://tex.stackexchange.com/questions/35116/specifying-proportional-numbers-when-using-libertine-with-pdflatex) \addfontfeature{Numbers=Proportional}

\title{Machine learning-based analysis of hyperspectral images for automated sepsis diagnosis}

\author{
\name{Maximilian Dietrich\textsuperscript{*1} \& Silvia Seidlitz\textsuperscript{*2,3}, Nicholas Schreck\textsuperscript{4}, Manuel Wiesenfarth\textsuperscript{4}, Patrick Godau\textsuperscript{2,3}, Minu Tizabi\textsuperscript{2}, Jan Sellner\textsuperscript{2,3}, Sebastian Marx\textsuperscript{1}, Samuel Kn\"odler\textsuperscript{5}, Michael M. Allers\textsuperscript{5}, Leonardo Ayala \textsuperscript{2,7}, Karsten Schmidt\textsuperscript{8}, Thorsten Brenner\textsuperscript{8}, Alexander Studier-Fischer\textsuperscript{5}, Felix Nickel\textsuperscript{5}, Beat P. M\"uller-Stich\textsuperscript{5}, Annette Kopp-Schneider\textsuperscript{4}, Markus A. Weigand\textsuperscript{1**} \& Lena Maier-Hein\textsuperscript{2,6,7**}}
\affil{
\footnotesize
\textsuperscript{*/**}authors contributed equally \\
\textsuperscript{1}Department of Anesthesiology, Heidelberg University Hospital, Heidelberg, Germany \\
\textsuperscript{2}Division of Computer Assisted Medical Interventions, German Cancer Research Center (DKFZ), Heidelberg, Germany \\
\textsuperscript{3}HIDSS4Health - Helmholtz Information and Data Science School for Health, Karlsruhe/Heidelberg, Germany \\
\textsuperscript{4}Division of Biostatistics, German Cancer Research Center (DKFZ), Heidelberg, Germany \\
\textsuperscript{5}Department of General, Visceral, and Transplantation Surgery, Heidelberg University Hospital, Heidelberg, Germany.\\
\textsuperscript{6}Faculty of Mathematics and Computer Science, Heidelberg University, Heidelberg, Germany \\
\textsuperscript{7}Medical Faculty, Heidelberg University, Heidelberg, Germany \\
\textsuperscript{8}Department of Anesthesiology and Intensive Care Medicine, University Hospital Essen, University Duisburg-Essen, Essen, Germany
}
}

\maketitle

\begin{abstract}
Sepsis is a leading cause of mortality and critical illness worldwide. While robust biomarkers for early diagnosis are still missing, recent work indicates that hyperspectral imaging (HSI) has the potential to overcome this bottleneck by monitoring microcirculatory alterations. Automated machine learning-based diagnosis of sepsis based on HSI data, however, has not been explored to date. Given this gap in the literature, we leveraged an existing data set to (1) investigate whether HSI-based automated diagnosis of sepsis is possible and (2) put forth a list of possible confounders relevant for HSI-based tissue classification. While we were able to classify sepsis with an accuracy of over \SI{98}{\%} using the existing data, our research also revealed several subject-, therapy- and imaging-related confounders that may lead to an overestimation of algorithm performance when not balanced across the patient groups. We conclude that further prospective studies, carefully designed with respect to these confounders, are necessary to confirm the preliminary results obtained in this study.

\end{abstract}

\section{Purpose}

Sepsis, a life-threatening organ dysfunction resulting from a dysregulated host response to infection \cite{singer_third_2016}, is a leading cause of mortality and critical illness worldwide with sepsis-related deaths representing \SI{19.7}{\percent} of deaths in 2017 \cite{rudd_global_2020}. One of the major challenges is the early detection of sepsis because irreversible organ damage increases the mortality rate with every hour the antimicrobial intervention is delayed \cite{ferrer_empiric_2014}.

Despite decades of clinical research, robust biomarkers for sepsis detection are still missing \cite{moor_early_2021}. Recent studies have investigated diagnosing and predicting sepsis from digital patient data including laboratory, vital, genetic and molecular data and health history \cite{fleuren_machine_2020, moor_early_2019, kaji_attention_2019, bloch_machine_2019}. Other authors have put forth the hypothesis that hyperspectral imaging (HSI) may be suitable for sepsis diagnosis as it potentially allows for the monitoring of microcirculatory alterations \cite{dietrich_bedside_2021, kazune_impact_2019, kazune_relationship_2019, saknite_novel_2017}.
%However, the potential of HSI to provide a robust, non-invasive and fast classification of sepsis patients has not yet been investigated. 
Specifically, a recent study~\cite{dietrich_bedside_2021} found characteristic patterns when comparing HSI data of the palm of the hand and thigh obtained from sepsis patients with HSI data from healthy subjects. It remains to be investigated, however, whether automated sepsis diagnosis based on such signatures is possible with machine learning.

The field of machine learning has rapidly gained importance for automated disease diagnosis in general. In fact, a large body of clinical success stories has been generated in the past few years, spanning various medical domains including dermatology~\cite{chan_machine_2020, liu_deep_2020, esteva_dermatologist-level_2017}, radiology~\cite{choy_current_2018, hosny_artificial_2018}, gastroenterology~\cite{sung_artificial_2020} and many more. However, an increasing number of papers also revealed severe flaws related to study designs, leading to biased algorithms and overestimation of performance (e.g. ~\cite{obermeyer_dissecting_2019, davis_calibration_2017, mac_namee_problem_2002}). Specifically, the issue of confounding bias has to date been given far too little attention. Confounding is \enquote{a \enquote{mixing} or \enquote{blurring} of effects. It occurs when an investigator tries to determine the effect of an exposure on the outcome, but then actually measures the effect of another factor, a confounding variable}~\cite{jager_confounding_2008}. Typically, confounding leads to an overestimation of algorithm performance, as demonstrated in several recent papers~\cite{zhao_training_2020, roberts_common_2021, badgeley_deep_2019, winkler_association_2019}. A prominent example was recently provided by Roberts et al.~\cite{roberts_common_2021}. They show that more than 15 papers, partially published in prestigious journals, merged two widely known data sets for training and testing a classifier to diagnose COVID-19 from images: A pneunomia data set comprising only infants between one and five, and a second data set comprising only adult COVID-19 patients. The authors concluded that a classifier trained on such merged data sets \enquote{is likely to overperform as it is merely detecting children versus adults}. Another paper in the context of radiological data science demonstrated an algorithm to detect hip fractures that primarily based its decisions on confounding variables~\cite{badgeley_deep_2019}. In contrast, we identified only one paper in the field of medical HSI that mentions potential limitations of findings due to confounding bias \cite{chin_evaluation_2011}. Overall, classification of medical HSI data is becoming increasingly popular, but the issue of confounding has not yet been systematically investigated. 

Given the gap in the literature with respect to both automated sepsis diagnosis based on HSI and confounder analysis related to medical HSI, the contribution of this work is twofold: Firstly, we show that machine learning-based sepsis diagnosis with high accuracy is possible on the available data set~\cite{dietrich_bedside_2021}. Secondly, we perform a systematic analysis of possible confounders in the data to estimate generalization capabilities of the algorithm developed.

\section{Materials and Methods}

The following sections present the available dataset (\autoref{sec:data}), our machine-learning based approach to automatically diagnosing sepsis (\autoref{sec:ML}) and our systematic analysis of potential confounders in the HSI data set (\autoref{sec:confounder}).

\subsection{Data set}
\label{sec:data}

Our analysis is based on the dataset described in \cite{dietrich_bedside_2020, dietrich_bedside_2021, dietrich_hyperspectral_2020}, which was acquired at the Heidelberg University Hospital after approval by the local Institutional Ethics Committee (DE/EKB03, study reference number: S-148/2019). The study is registered with the German Clinical Trials Register (DRKS00017313). The dataset comprises HSI data of palm of the hand and thigh skin for 25 healthy subjects, 25 patients with sepsis or septic shock and 25 patients undergoing pancreatic surgery. All subjects were adults. Sepsis patients were included on admission to the surgical intensive care unit (ICU) if all Sepsis-3 criteria were met \cite{singer_third_2016} and the onset of the sepsis was less than \SI{24}{h} ago. Healthy subjects were included if they had no acute or chronic disease. Patients with an elective pancreatic resection surgery were included preoperatively if an open surgery and a postoperative ICU admission were planned.

\begin{figure}[H]
\centering
\includegraphics[width=\textwidth]{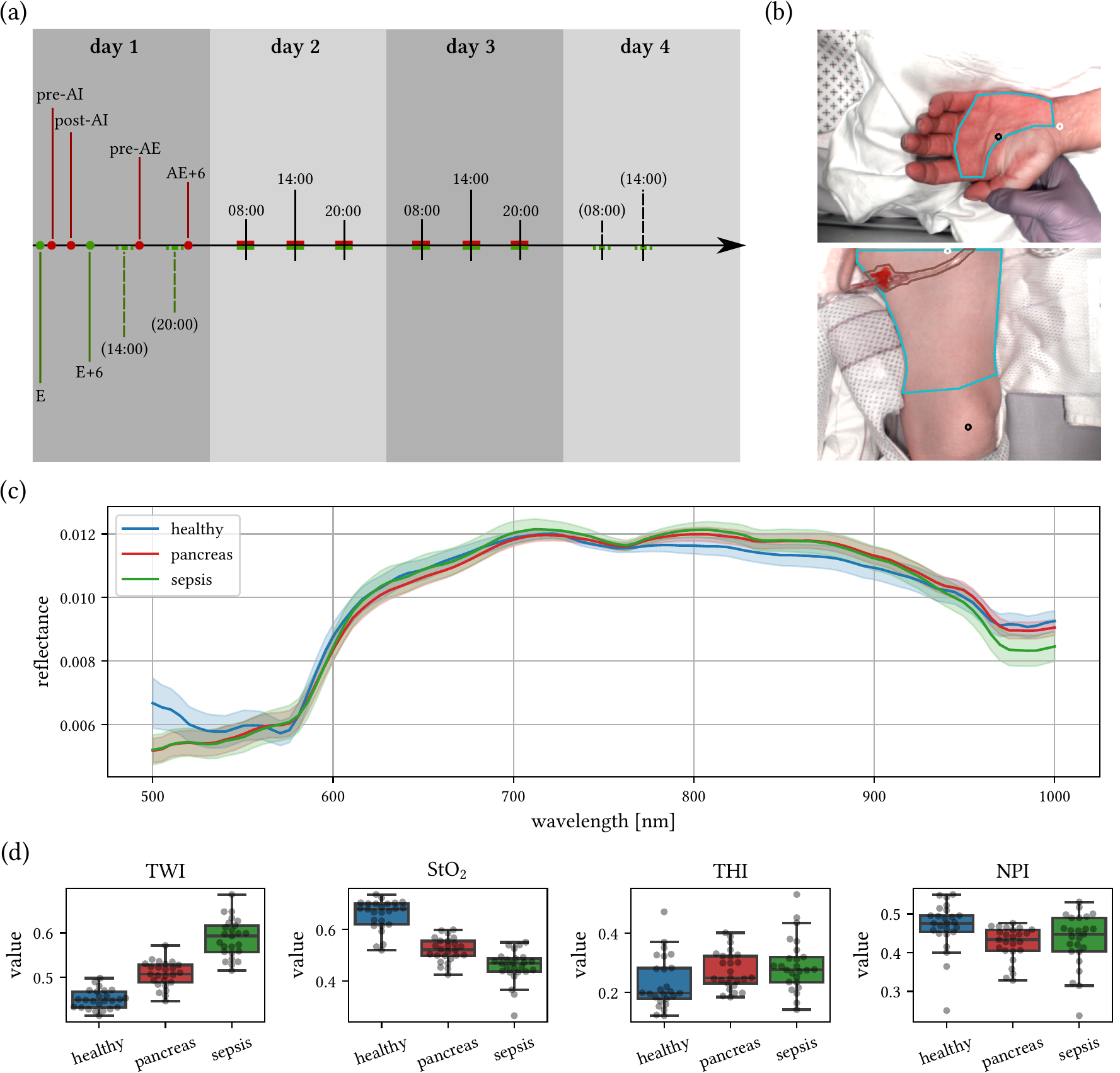} 
\caption{Data set characteristics (a) Overview of scheduled acquisition time points for patients undergoing pancreatic surgery (red) and sepsis patients (green). Dashed lines and brackets indicate optional time points that were measured only in the case of an either early or very late admission. (b) Annotations for an exemplary hand (top) and thigh (bottom) image: lines indicate annotated skin area (cyan), orientation of body part (black and white circle) and obscured skin area (brown). (c) Mean (solid line) and standard deviation area (shaded) computed on patient-wise aggregated median spectra available for the three subject groups healthy (blue), pancreatic surgery (red) and sepsis (green). (d) Distribution of the patient-wise aggregated median tissue parameters for the three different subject groups.}
\label{fig:data}
\end{figure}

While HSI data of healthy subjects were acquired on a single time point, sepsis patients and patients undergoing pancreatic surgery were imaged at defined time points during an approximately \SI{72}{h} observation period. For patients undergoing pancreatic surgery, the acquisition time points were before induction of anesthesia (pre-AI), after induction of anesthesia (post-AI), before anesthesia emergence (pre-AE), approximately 6 hours postemergence of anesthesia (AE+6) and three times daily (08:00, 14:00, 20:00 ± \SI{2}{h}) for the first two postoperative days, yielding 10 consecutive measurements per patient. For sepsis patients, the intended acquisition time points were on admission to the ICU (E), approximately 6 h afterwards (E+6) and in the case of an admission at an early time of day, up to two additional measurements were taken in the afternoon and evening (14:00, 20:00 ± \SI{2}{h}). On the subsequent 2--3 days, HSI data was usually acquired  three times a day (08:00, 14:00, 20:00 ± \SI{2}{h}). An illustration of the scheduled acquisition time points for both patient groups is given in \autoref{fig:data}\,(a). For most of the sepsis patients, measurements are available for 9 distinct time points. However, it should be noted that 5 out of the 25 sepsis patients died and one got transferred to another hospital within the \SI{72}{h} observation period. For three patients, divergence from the measurement scheme was required due to them undergoing a surgery. Furthermore, measurement of the thigh region was not feasible for all available time points as in 17 cases patients were in prone positions.

In addition to HSI data, clinical data was documented, including:
\begin{itemize}
    \item patient demographics (age, sex, weight, height, body mass index (BMI) at inclusion), 
    \item hemodynamic monitoring (cardiac index, cardiac output, heart frequency, systolic, diastolic and mean arterial blood pressure (MAP), central venous pressure, global end-diastolic volume, systemic vascular resistance, extravascular lung water index, stroke volume, stroke volume variation, pulse pressure variation and diameter and velocity time integral of left ventricular outflow tract),
    \item routine clinical laboratory parameters (concentrations of albumin, hemoglobin (Hb), thrombocytes, creatinine, urea and bilirubin, hematocrit and prothrombin time) and inflammatory markers (concentrations of leukocytes, C-reactive protein (CRP) and procalcitonin, body temperature),
    \item blood gas analysis parameters (pH, oxygen partial pressure, partial pressure of carbon dioxide, bicarbonate concentration, base excess, lactate concentration),
    \item peripheral oxygen saturation ($\mathrm{spO_2}$),
    \item vasopressor support (doses of arterenol, epinephrin, vasopressin and dobutrex and vasoactive-inotropic score (VIS)),
    \item fluid balance (fluid input separated according to kind (crystalloid, colloid, erythrocyte concentrate, fresh frozen plasma), fluid output (including blood loss during surgery), computed balance),
    \item ventilation parameters (fraction of inspired oxygen, positive end-expiratory pressure),
    \item 30-day mortality and daily measurements of prognostic markers (SOFA score, ASA score and Syndecan-1),
    \item detailed diagnosis and suspected focus of infection,
    \item approximate time point and duration of surgical intervention and kind of procedure.
\end{itemize}

More details on which parameters are available for which subject groups and acquisition time points are available from \cite{dietrich_bedside_2020}.

The HSI data was acquired with the hyperspectral camera system TIVITA\textsuperscript{\textregistered} Tissue (Diaspective Vision GmbH, Am Salzhaff, Germany), which is an approved medical Class I device. It captures hyperspectral images with a size of $640\times480$ pixels (width $\times$ height) and a spectral resolution of approximately \SI{5}{nm} in the spectral range between \SI{500}{nm} and \SI{1000}{nm}, resulting in a 100-dimensional reflectance spectrum for each pixel. The camera system captures an area of approximately 20 $\times$ \SI{30}{cm} at an imaging distance of about \SI{50}{cm}. To ensure uniform illumination of the scene, the room light was dimmed during image acquisition and the integrated halogen lighting unit and stray light sensor of the camera system were used \cite{dietrich_bedside_2021}. More technical details are available in \cite{kulcke_compact_2018}.

Skin areas were annotated according to a detailed annotation protocol designed for this study. For hand images, the palm excluding the thenar was annotated, whereas for thigh images, the entire visible thigh skin above the patella was annotated. Two examplary annotations are shown in \autoref{fig:data}\,(b). Shaded and obscured areas, wounds and pressure marks were excluded from the analysis. For each image, the median over all pixel spectra comprised in the annotation was computed, in the following referred to as median spectrum. \autoref{fig:data}\,(c) shows the distributions of the median spectra for the three different subject groups. Additionally, the camera system provides estimations of tissue parameters such as oxygenation ($\mathrm{StO_2}$), perfusion (NIR perfusion index, NPI), Hb content (tissue hemoglobin index, THI) and water content (tissue water index, TWI). These are computed from the hyperspectral images as decribed in \cite{holmer_hyperspectral_2018}. For each image, the median over the tissue parameters inside the annotated area is computed. \autoref{fig:data}\,(d) shows the distributions of the median tissue parameters for the three different subject groups.

\subsection{ML-based sepsis classification}
\label{sec:ML}

Linear Discriminant Analysis (LDA) is a classical statistical method which discriminates two or more classes of objects by finding the set of projection vectors which maximize the ratio of between-class scatter against within-class scatter \cite{li_lda_2009}. Here, we train an LDA classifier to discriminate between sepsis and non-sepsis subjects based on the median spectra of each annotated region, as described in Sec. \ref{sec:data}. To quantify the classification performance, we compiled a large number of random train-test-splits via bootstrap subsampling in order to overcome a potential bias resulting from a single small test set. More specifically, we generated 500,000 random train-test-splits with each test set comprising 5 healthy, 5 pancreatic surgery and 5 sepsis patients and the training set comprising the remaining 60 patients. On each training set an LDA classifier was fitted (without touching the corresponding test set), resulting in 500,000 different LDA classifiers. To account for the different numbers of measurements performed per patient, metric values were always aggregated for all images of one patient before computing descriptive statistics. Reported mean values were averaged over the 500,000 test sets.

We address the following research questions:

\begin{enumerate}
\item Is there an advantage in using median spectra over tissue parameter images?
\item How does classification performance change with different choices of measurement site (only hand versus only thigh versus using both sites for training and testing)?
\item What is the influence of the choice of measurement time point on the classification performance? How well does the algorithm discriminate the sepsis from non-sepsis subjects at inclusion?
\end{enumerate}

\subsection{Confounder analysis in sepsis classification}
\label{sec:confounder}

To investigate potential biases related to our classifier, we generated a list of possible confounders related to our HSI data set. These can be grouped in 3 categories: 

\paragraph*{Subject-related confounders} refer to variables that measure the health status and demographics of a subject. Here, we distinguish between static and dynamic parameters: 

\textit{Static parameters} are only acquired once at inclusion and can be assumed to be approximately constant during the measurement period. These include age, sex and body mass index (BMI) as well as variables related to the medical history such as comorbidities. % rahrovan et al: review of studies of skin color differences between male and female, 

\textit{Dynamic parameters} vary during the measurement period. These include, for example, the MAP as a measure for hemodynamic stability, the $\mathrm{spO_2}$, Hb concentration and bilirubin concentration. Changes in $\mathrm{spO_2}$, Hb and bilirubin concentrations can be a consequence of sepsis or a pancreatic surgery but such effects are either rare \cite{ansari_hemorrhage_2017} or are addressed in therapy \cite{jung_relationship_2019} and therefore are not mediators in the causal pathway. They are potentially important confounders since bilirubin, oxyhemoglobin and deoxyhemoglobin are biological chromophores that shape the measured reflectance spectra \cite{beschastnov_current_2018, kudavelly_simple_2011}.

\paragraph*{Therapy-related confounders} refer to potential systemic changes resulting from the treatment of a patient and can be induced by medication and surgical trauma, for example.  
%Generally, the therapy of pancreatic surgery and sepsis patients was determined by the care-giving physician according to local protocols. 
Both sepsis patients as well as intraoperative and postoperative patients receive intravenous fluid therapy and cardiovascular support. The corresponding variables are the fluid balance (difference between fluid input and output) and the VIS. These have a potential effect on HSI measurements. Due to capillary leakage, fluid therapy can exacerbate edema formation in sepsis patients \cite{marx_fluid_2003, malbrain_principles_2018}. The increased tissue water content affects the measured absorption spectra since water is one of the main biological absorbers in the near-infrared spectral region \cite{cao_multispectral_2013}. Cardiovascular support creates vasoconstriction or increases cardiac contractility which can alter perfusion \cite{annane_global_2018, vanvalkinburgh_inotropes_2021}. Several works have shown that changes in tissue perfusion can be detected from HSI data \cite{holmer_ability_2017, chin_hyperspectral_2012, zuzak_visible_2002}, thus making it an important potential confounder. Another potential therapy-related confounder is the performance of a major surgery which can be followed by generalized edema \cite{vaughan-shaw_oedema_2013} affecting the near-infrared region of the spectra as described above. 
%Also sedation and ventilation should be considered as potential confounders.

\paragraph*{Imaging-related confounders} arise from changes in hardware, software or measurement protocols throughout the data acquisition process. Examples are different imaging devices, different calibrations applied to a given device, differences in the pose of the camera relative to the acquisition subject and differences in illumination or temperature resulting, for example, from acquisitions in different rooms. All of these have a direct potential effect on the measurements but we are not aware of work that has systematically investigated the effect of such parameters on HSI measurements.

\autoref{sec:confounder-analysis} provides descriptive statistics about the potential confounders identified, separated by patient group.

%We aim to address the potential confounders in the following way:

%\begin{enumerate}
%    \item give an introduction to LMMs
%    \item show equation and explain choice of fixed and random effects
%    \item show results from LMMs [TODO discuss with statisticians what else could be shown apart from explained variance]
%    \item discuss implications for classification
%\end{enumerate}

\section{Results}
\label{sec:results}

\subsection{ML-based sepsis classification} 

\begin{figure}[H]
\centering
\includegraphics[width=\textwidth]{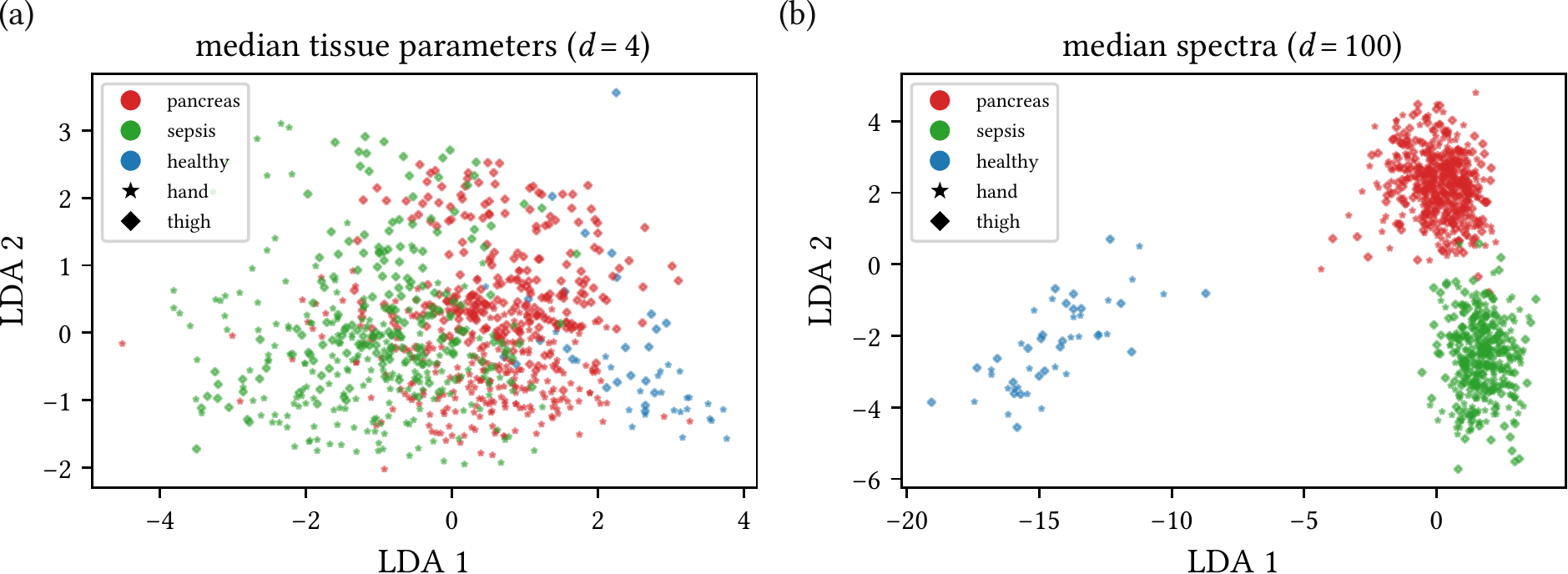}
\caption{LDA projections obtained when fitting a classifier on the entire data set using  (a) the median tissue parameters (dimensionality $d = 4$) or (b) the median spectra ($d = 100$). Each scatter point represents one image of a sepsis (green), pancreatic surgery (red) and healthy (blue) subject. The symbol encodes the measurement site.}
\label{fig:LDA-plots}
\end{figure}

\paragraph*{Effect of input features} While LDA failed to provide a good separation of the three subject groups based on the median parameter values (\autoref{fig:LDA-plots}\,(a)), the median skin spectra for healthy, pancreatic surgery and sepsis subjects showed distinct clusters when projecting on the first two components of LDA (\autoref{fig:LDA-plots}\,(b)). Therefore median spectra were used for all subsequent analyses. In the following, the classification performance is compared for different measurement sites and time points. For all settings, the same 500,000 train-test-splits were used.

\begin{figure}[H]
\centering
\includegraphics[width=\textwidth]{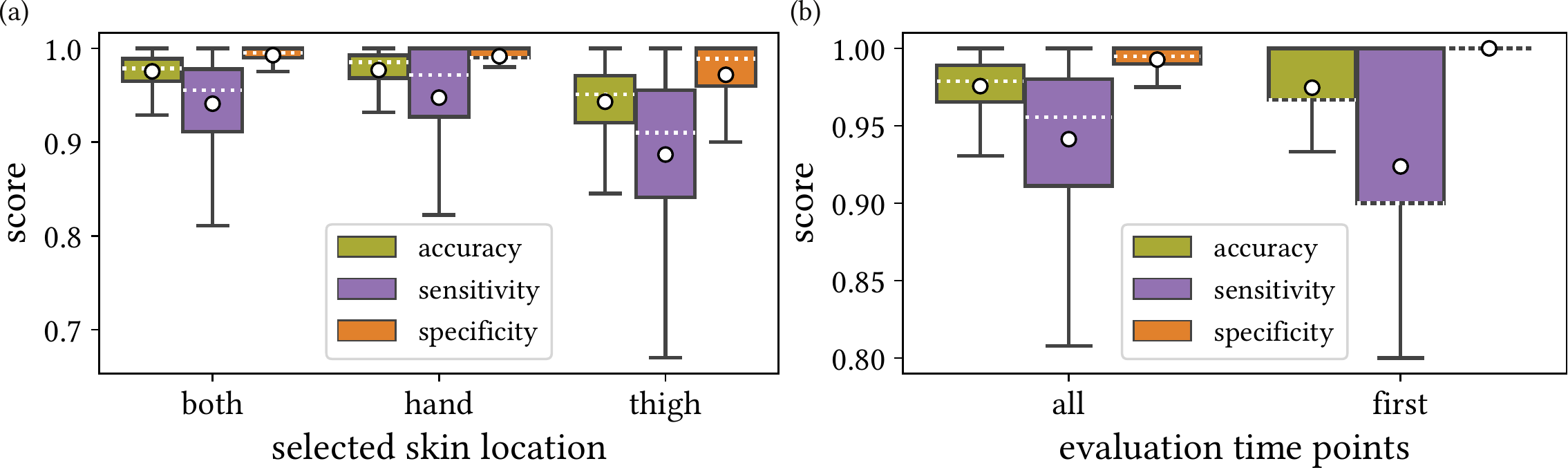}
\caption{ML performance of the LDA classifier itemized by (a) different locations, (b) aggregating all or only the first time point. Each boxplot shows the quartiles of the distribution with the whiskers extending to 1.5 times the interquartile range, the median as white dotted line and the mean as white circle.}
% , change over time for (c) sepsis and (d) pancreatic surgery patients. Each boxplot shows the quartiles of the distribution with the whiskers extending to 1.5 times the interquartile range, the median as white dotted line and the mean as white circle.
%\caption{ML classification performance. (a) LDA projection obtained when fitting a classifier on the entire data set. Each scatter point represents one image of a sepsis (green), pancreatic surgery (red) and healthy (blue) subject. (b) Boxplot illustrating the distribution of classification accuracies obtained for the 500,000 randomly drawn test sets. The box shows the quartiles of the distribution, the whiskers extend up to 1.5 times the interquartile range and the horizontal black line indicates the median value.}
\label{fig:ML-results}
\end{figure}

\paragraph*{Effect of measurement site} For comparing different measurement sites (\textit{hand}, \textit{thigh} and the combination of \textit{both} sites), all available time points were included in training and testing of the classifier. While similar accuracy, sensitivity and specificity was obtained in the case of classifiers trained and tested on \textit{both} and \textit{hand} measurement sites, a drop in performance was observed for the \textit{thigh} skin spectra (\autoref{fig:ML-results}\,(a)).

\paragraph*{Effect of time point} While inclusion of all available time points describes the ability to classify sepsis at any stage of disease and therapy within the 72 hours observation period, another clinical target is the classification at inclusion. \autoref{fig:ML-results}\,(b) shows accuracy, sensitivity and specificity when testing on \textit{all} time points compared to testing on the \textit{first} time points (admission to ICU for sepsis patients, before anesthetic induction for pancreatic surgery patients and inclusion for healthy subjects). In both cases, training was performed on all available time points and both measurement sites were included. The classification performance is similar with a mean accuracy of 0.98, sensitivity of 0.94 and specificity of 0.99 for testing on all time points, and a mean accuracy of 0.98, sensitivity of 0.92 and specificity of 1.00 for testing on the first time point. %\autoref{fig:ML-results}\,(c) gives a more detailed insight into the sensitivities for different time points for sepsis patients and \autoref{fig:ML-results}\,(d) shows the specificities for different time points for pancreatic surgery patients. It should be noted that the significance of \autoref{fig:ML-results}\,(c) is limited due to fluctuations in the measurement scheme (e.g. patients dying within the \SI{72}{h} observation period observation period). 

%For each of the 500,000 test sets, the accuracy of the LDA predictions is computed, yielding an overall accuracy of 0.97 ($s = 0.02$) (see Fig. \ref{fig:ML-results}\,(b)). \autoref{fig:ML-results}\,(a) provides a visual impression of the good linear discriminability of median skin spectra for healthy subjects, pancreatic surgery and sepsis patients by projecting on the first two components of LDA.

\subsection{Confounder analysis}
\label{sec:confounder-analysis}

\begin{figure}[H]
\centering
\includegraphics[width=\textwidth]{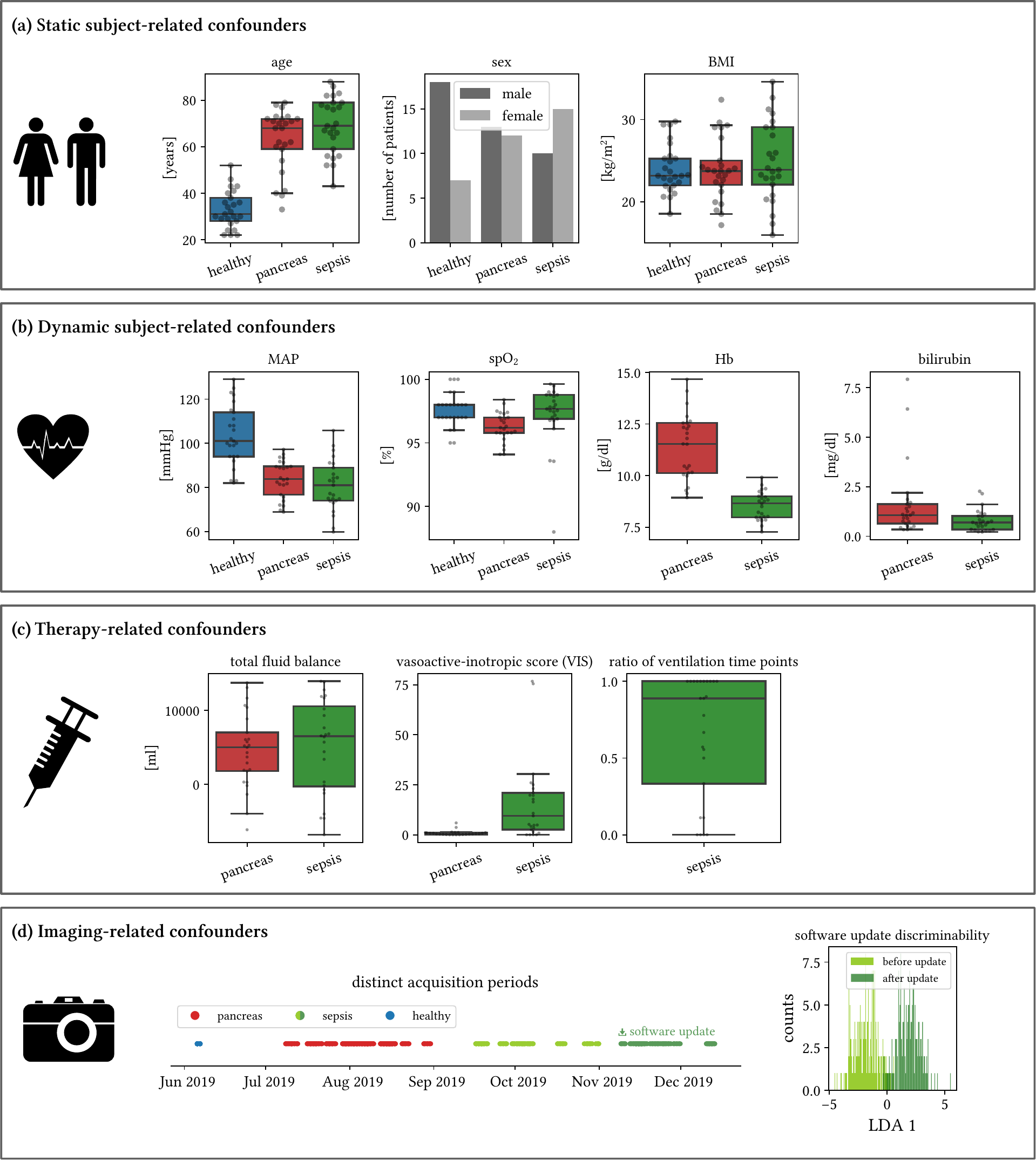} 
\caption{Selection of potential confounders. Parameters were averaged for each subject before generating box and swarm plots. Boxes show the quartiles of the distribution, the whiskers extend to 1.5 times the interquartile range and the horizontal black line indicates the median value. (a) \textit{Static subject-related confounders}: Distributions are shown for the parameters age, sex and BMI. (b) \textit{Dynamic subject-related confounders}: Distributions are shown for the parameters MAP, $\mathrm{spO_2}$, Hb concentration and bilirubin concentration. Since measurements of Hb and bilirubin concentrations are not available for the healthy patients, only pancreatic surgery and sepsis groups are shown in these cases. (c) \textit{Therapy-related confounders}: Distributions are shown for the fluid balance, VIS as a measure of cardiovascular support, and ratio of time points at which artificial ventilation was recquired for sepsis patients. Only patient groups who received the respective therapy are shown. (d) \textit{Imaging-related confounders}: Visualization of the distinct acquisition periods for the three subject groups (left). Each scatter point corresponds to one acquired image. A software update was performed in November 2019 as indicated by a change in the color tone. Histograms of LDA projections obtained when fitting a classifier to all median spectra of sepsis patients (right). Color tone according to software version used at acquisition.}
\label{fig:confounders}
\end{figure}

Distributions of those potential confounding variables for which measurements are available are provided in \autoref{fig:confounders}. 

\paragraph*{Subject-related confounders:} Distributions of the static subject-related confounders are illustrated in \autoref{fig:confounders}\,(a). Especially the shift in age distribution between healthy volunteers and sepsis/pancreatic surgery patients is a potentially strong confounder. For sepsis patients and healthy volunteers, no information on comorbidities is available. 
%For patients undergoing pancreatic surgery, generally, pancreatic cancer is the cause for the partial or total pancreatectomy. An effect of pancreatic cancer on the measured spectra cannot be excluded.
%\paragraph*{Subject-related confounders:}  
Distributions of the dynamic subject-related confounders are illustrated in \autoref{fig:confounders}\,(b). While the MAP, $\mathrm{spO_2}$ and bilirubin concentration match rather well for at least one sepsis and non-sepsis group, a distributional shift is present for the blood Hb concentration between non-sepsis and sepsis patients. Hb concentration is thus considered as a potentially strong confounder.
%Body temperature is not considered a confounding variable since it would rather be a mediator in the causal pathway (inflammation in sepsis causing increase in body temperature potentially causing a certain change in the measured spectra).

\paragraph*{Therapy-related confounders:} While all pancreatic cancer patients underwent surgery between the observation time points post-AI and pre-AE, only 3 out of 25 sepsis patients had surgery performed during the measurement period. Potential surgeries before inclusion of sepsis patients in the study had not been documented (sepsis could be a postoperative complication). Sedation and ventilation should be considered as further potential confounders. While patients undergoing pancreatic surgery were only sedated and ventilated intraoperatively, 19 out of 25 sepsis patients were sedated and ventilated at least once during the 72 hours measurement period and \SI{40}{\%} of the sepsis patients required permanent sedation and ventilation. The ratio of time points at which ventilation was required for the sepsis patients is illustrated in \autoref{fig:confounders}\,(c) together with the fluid balance and VIS distributions.

\paragraph*{Imaging-related confounders:} A change in hardware did not occur in the course of the study that our data set is based on, but a software update was made. While the images of all healthy and pancreatic surgery subjects were acquired prior to the software update, for \SI{48}{\%} of the sepsis patients images were taken after the software update. Median spectra of sepsis patients acquired prior to and after the software update can be separated by use of LDA (see \autoref{fig:confounders}\,(d)). Since images for the three different subject groups were not acquired in parallel but at different seasons throughout the year (\autoref{fig:confounders}\,(d)), a seasonal effect is considered to be a potential confounder. All subjects are of a comparable light skin type but the content of melanin, which is an important skin chromophore, is known to be elevated following summer compared to following winter \cite{hexsel_variation_2013}. However, this effect is likely very small in palms and thighs, since these skin areas have little UV exposure. Another potential confounder is the systematic difference in hand postures. While hands of sepsis patients and sedated pancreatic surgery patients tend to be clenched, hands of healthy subjects and non-sedated pancreatic surgery patients tend to be open. A systematic change in camera perspective is not apparent. Changes in illumination are documented with the stray light detection algorithm of the camera system and affect only a small number of images from both pancreatic surgery and sepsis patients.

\section{Discussion}

To our knowledge, we are the first to present a machine-learning based approach for automated sepsis diagnosis based on HSI. While we achieved outstanding performance on a previously published data set, our analysis also identified a set of confounders that may have led to an overestimation of algorithm performance. In ongoing statistical analyses, we are now studying to which extend these contribute to the measured spectra. Additionally, independent studies in which imaging-related confounders such as hand postures and camera perspectives are systematically analyzed, are performed.

In the present study, we used LDA as a relatively simple traditional machine learning method, and it yielded outstanding performance on our data. If the method turns out to be insufficient in future studies with matched confounders, more sophisticated methods, such as deep learning or random forests, which we applied successfully to spectral images in previous work \cite{moccia_uncertainty-aware_2018, wirkert_robust_2016, grohl_learned_2021, grohl_semantic_2021}, can be explored.

The key strength of HSI systems for sepsis diagnosis is their fast and non-invasive assessment which does not require complex clinical infrastructure and could thus be utilised in a mobile manner. This is a huge advantage over sepsis diagnosis based on digital patient data, which generally requires access to laboratory and clinical data and is thus limited to patients already admitted to an ICU. However, a disadvantage of HSI-based sepsis diagnosis is that it needs to be actively recorded by the medical staff and is thus more labor-intensive than the processing of automatically recorded digital patient data. A key strength of algorithms based on digital patient data is that they have been successfully applied to the task of early prediction of sepsis up to several hours before the actual onset (e.g. \cite{moor_early_2019}). 

It is worth mentioning that the potential of HSI-based sepsis diagnosis for early diagnosis of sepsis could not be studied with the available data and should thus be considered in the design of future studies. We further plan to extend the existing data in a prospective study that accounts for the identified confounders, aims for the early inclusion of patients prior to sepsis onset and inclusion of border cases, and features a more comprehensive sample size. By combining the acquisition of HSI data with the continuous recording of digital patient data, we aim to compare the sepsis diagnosis capabilities of both data sources and investigate potential synergies.

In conclusion, this work shows that highly accurate sepsis diagnosis may be possible from HSI data. Further studies are required to confirm this hypothesis.

\section*{Acknowledgements}
This project has received funding from the European Research Council (ERC) under the European Union’s Horizon 2020 research and innovation program (grant agreement No. [101002198]; NEURAL SPICING). The present contribution is further supported by the Helmholtz Association under the joint research school \enquote{HIDSS4Health – Helmholtz Information and Data Science School for Health} and the Helmholtz Imaging Platform (HIP).

\section*{Compliance with ethical standards}
The authors declare that they have no conflict of interest.

\bibliographystyle{ieeetr} %TODO other style apacite?
\bibliography{references.bib}

\end{document}